\documentclass[seceq]{ptptex}

\usepackage{graphicx}

\newcommand{\be}{\begin{equation}}
\newcommand{\ee}{\end{equation}}
\newcommand{\bea}{\begin{eqnarray}}
\newcommand{\eea}{\end{eqnarray}}
\newcommand{\ba}{\begin{array}}
\newcommand{\ea}{\end{array}}

\newcommand{\p}{\partial}

\title{
Gravitational string-membrane hedgehog and internal structure of black holes%
}

\author{
Hikaru \textsc{Kawai}$^{1,2,}$\footnote{E-mail: hkawai@gauge.scphys.kyoto-u.ac.jp} 
and Toshihiro \textsc{Matsuo}$^{3,}$\footnote{E-mail: tmatsuo@yukawa.kyoto-u.ac.jp}
}

\inst{
$^1$Department of Physics, Kyoto University, Kyoto 606-8502, Japan
\\
$^2$Theoretical Physics Laboratory, Nishina Center, RIKEN, Wako, Saitama, 351-0198, Japan
\\
$^3$Okayama Institute for Quantum Physics, Kyoyama 1-9-1, Okayama 700-0015, Japan
}

\abst{
We investigate charged Nambu-Goto strings/membrane systems in the Einstein-Maxwell theory in $3+1$ dimensions. We first construct a charged string hedgehog solution that has a single horizon and conical singularity. Then we examine a charged membrane system, and give a simple derivation of its self-energy. We find that the membrane may form an extremal Reissner-Nordstr\"om black hole, but its interior is a flat spacetime. Finally by combining the charged strings and the membrane we construct black hole solutions that have no singularities inside the horizons. We study them in detail by varying the magnitude of the two parameters, namely, the charge times the membrane tension and the string tension. We also argue that the strings have, due to the large redshift inside the system, a fair amount of degrees of freedom that may explain the entropy of the corresponding black holes.
}

\begin{document}

\maketitle

\section{Introduction}
The Reissner-Nordst\"om solution is the unique, asymptotically flat, static and spherically symmetric solution of the source-free Einstein-Maxwell theory in $3+1$ dimensions, and gives the second simplest class of black holes.
The existence of the electric field makes the geometry qualitatively different from that of the Schwarzschild black hole. In fact, the solution has two distinct horizons, one is outer event horizon and the other is inner Cauchy horizon. 

Strings may be regarded as extremal forms of electromagnetic fields, that is, infinitely thin flux tube. Unlike the electromagnetic fields, however, strings have no transverse stresses though they have the same equation of states as that of the electromagnetic fields in the direction they extend. 
Ensembles of such strings form ``string matter" that is intrinsically anisotropic.
We are interested in constructing black holes by making use of the string matter and investigate the internal structure of the solutions. Since the strings have tension and tend to shrink, the string matter should have charges in order to stabilize the configurations. 
Thus the solutions inevitably provide the Reissner-Nordst\"om solution outside the matter, but the internal structure, that is our main concern, turns out to be different. 
In addition to the strings and charges, we will introduce a spherical membrane that is self-gravitating.
A spherically symmetric membrane has no singularity, because its interior is flat spacetime.
Furthermore, strings can be attached to membranes, and we have more intriguing solutions if we combine them.
Thus we can make various static solutions that have the same metric as the Reissner-Nordst\"om solution outside the outer horizon but have different internal structures.
The existence of these solutions may shed some light, though not conclusive, on the problem of the origin of the black hole entropy.


The organization of this paper is as follows.
In section 2 we investigate systems consisting of strings, membranes and charges in the Einstein-Maxwell theory in $3+1$ dimensions.
We first introduce a certain charged stringy configuration which has a conical singularity at the origin. We find an event horizon if the string tension becomes larger than a critical value.
Next we study a charged spherical membrane that may approach an extremal black hole, though the interior is a flat spacetime.
Then we combine these two solutions to make configurations that have horizons but no singularity.
Section 3 is devoted to brief discussions on the entropy of the models by considering random walk model of strings. 

Throughout this paper we use units in which $G=1$ and $c=1$.

\section{Stringy hedgehog configurations}
We investigate static and spherically symmetric configurations in the Einstein-Maxwell theory with sources consisting of strings, membranes and electric charges.
In the Schwarzschild coordinate system, the metric is given by
\bea
ds^2=-e^{2\phi(r)}dt^2+h(r)dr^2+r^2(d\theta^2+\sin^2\theta d\varphi^2) .
\eea
Since we consider hedgehog like configurations and do not assume local isotropy,   
the energy-momentum tensor is written in the form ,
\bea
{T^\mu}_\nu =
\left(
\begin{array}{cccc}
-\rho(r) & 0 & 0  &0 \\0 & P_r(r) & 0 & 0 \\0 & 0 & P_\perp(r) &0 \\0 & 0 & 0 & P_\perp(r)
\end{array}
\right) ,
\eea
where $\rho, P_r, P_\perp$ are the energy density, pressures in the radial and perpendicular directions, respectively.
The Einstein-Maxwell equations are 
\bea
R_{\mu \nu}-{1\over2}g_{\mu\nu}R=-8\pi T_{\mu \nu} ,
\label{Einstein}
\eea
and
\bea
{1\over \sqrt{-g}}\p_\mu\left(\sqrt{-g} g^{\mu \nu} g^{\rho \sigma} F_{\nu\sigma} \right) = J^\rho ,
\eea
where $J^\rho$ is the electric current, and $T_{\mu\nu}$ is the energy-momentum tensor given by
\bea
T_{\mu\nu}=T^{Maxwell}_{\mu\nu}+T^{matter}_{\mu\nu} .
\eea
Here $T^{Maxwell}$ is the contribution of the Maxwell field,
\bea
T^{Maxwell}_{\mu\nu}=-F_{\mu\rho}{F_{\nu}}^\rho+{1\over 4} g_{\mu\nu} F_{\rho\sigma}F^{\rho\sigma} ,
\eea
and $T^{matter}_{\mu\nu}$ is that of the matter whose concrete form is determined by the source we choose.

Here we assume that the charges are distributed on a thin spherical shell.  
Then from the symmetry and the equation of motion in vacuum, we have
the field strength outside the shell as
\bea
F_{0r}={q \over 4\pi r^2} ,
\eea
where $q$ is the total charge of the shell, and all the other components vanish.
There is no field strength inside the shell.
Then the energy density and pressure outside the shell are given by
\bea
\rho^{Maxwell}(r)= {Q^2 \over 8\pi r^4} , 
\quad
P_r^{Maxwell}(r)=- {Q^2 \over 8\pi r^4} , 
\quad
P_\perp^{Maxwell}(r)= {Q^2 \over 8\pi r^4} , 
\label{MaxwellEMT}
\eea
where $Q^2=q^2/(4\pi)$.

The Einstein equation \eqref{Einstein} reads
\bea
8\pi \rho&=&{1\over r^{2}}{d \over dr} r(1-h^{-1})  ,
\label{rhoeq}
\\
8\pi P_r&=&{2\phi' \over r h}-{1 \over r^2}(1-h^{-1}) ,
\label{Peq}
\\
8\pi P_\perp&=&{\phi'' \over h}+\left(\phi'-{h' \over 2h} +{1\over r}\right){\phi' \over h}
-{h' \over rh^2},
\label{Qeq}
\eea
where the prime denotes the derivative with respect to $r$.
We rearrange these equations so that physical meanings are transparent.
We first integrate the first equation \eqref{rhoeq} to have
\bea
h(r) = \left[1- {2m(r) \over r}
\right]^{-1} ,
\label{h}
\eea
where
\bea
m(r) = \int_0^r dr' 4\pi r'^2 \rho(r') 
\label{m}
\eea
is the mass inside the sphere of radius $r$.
The ADM mass $m_\infty $ is defined by $m(r)$ evaluated at $r=\infty$,
\bea
m_{\infty} = \lim_{r\to\infty}m(r) .
\eea
Substituting \eqref{m} into the second equation \eqref{Peq}, we find the gradient of the gravitational potential,
\bea
{d \phi \over dr} ={h \over r^{2}} \left( m+4 \pi r^3P_r \right) .
\label{dphidr}
\eea
Finally, we differentiate the second equation \eqref{Peq}, subtract the third equation \eqref{Qeq} multiplied by $2/r$ to eliminate the term $\phi''$, and add the second equation \eqref{Peq} multiplied by $2/r$ to get 
\bea
{dP_r \over dr} + {2 \over r} (P_r-P_\perp) 
+{h \over r^{2}} \left( m+4 \pi r^3P_r \right)(\rho+P_r) = 0 .
\label{balance}
\eea
This is the equation of equilibrium for anisotropic matter \cite{Lemaitre}. 
In fact, it shows that the gravitational force is balanced with the gradient of the pressure plus the difference of the pressures in the radial and perpendicular directions. Here the anisotropy of the source is responsible for the difference.\footnote{When $P_r=P_\perp$, the equation \eqref{balance} becomes the Tolman-Oppenheimer-Volkoff equation \cite{Oppenheimer:1939ne,Tolman:1939jz} for isotropic fluids.} 

\subsection{String hedgehog}
\label{String hedgehog}
As is explained in the introduction, we are interested in constructing the solutions by making use of strings that may be seen as infinitely thin electric flux tubes and have the same equation of states as the electromagnetic field in the direction they extend; on the other hand there is no transverse tension, as we will see shortly. Interestingly enough, it is possible to construct a charged black hole solution that has a single horizon.

Let us consider a static configuration that is composed of $N$ open strings with tension $\mu$. We assume that one end of each string is joined together at a point $O$, while the other end can move freely.
We attach an electric charge $e$ at the free end of each string, and the total charge amounts to $q=Ne$.
The charged end points tend to spread out because of electric repulsion, but they are pulled back by the strings. 
Therefore in large $N$ limit the strings extend radially and the end points distribute uniformly on a sphere of radius $r_1$ so that the whole configuration becomes spherically symmetric. We call this configuration ``charged string hedgehog".\footnote{For neutral stringy hedgehogs, see Refs. \citen{Guendelman:1991qb,Delice:2003zp}.} 
The region outside is a vacuum with electric field and the geometry is described by the Reissner-Nordstr\"om solution.
Note that there is a singularity (which turns out to be conical) at the center $O$.
We will remove it later, but in this subsection we leave it as it is. 
We also mention that we do not introduce any point mass or charge at the center.

\begin{figure}[h]
\begin{center}
\resizebox{5cm}{5.5cm}{\includegraphics{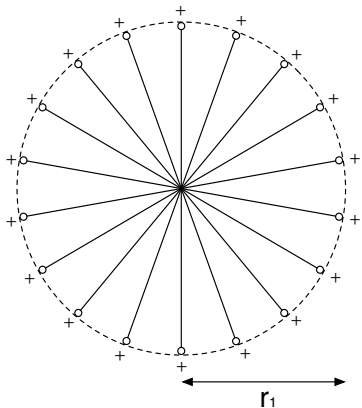}}
\end{center}
\caption{The charged string hedgehog. Two dimensional slice.}
\label{fig:hedgehog}
\end{figure}

The energy-momentum tensor of the $N$ strings is given by \eqref{emtNstrings} in Appendix \ref{emtstring}. Adding the contribution from the Maxwell field \eqref{MaxwellEMT}, we have
\bea
\rho(r) &=& {\sigma \over 4 \pi r^2} \theta(r_1-r) + {Q^2 \over 8 \pi r^4}\theta(r-r_1) ,
\label{rhoHH}
\\P_r(r)&=&-\rho(r) ,
\label{prHH}
\\P_\perp(r)&=&{Q^2 \over 8 \pi r^4}\theta(r-r_1) ,
\label{pperpHH}
\eea
where $\sigma = N \mu$. 
It follows from \eqref{m} that 
\bea
m(r)
&=&
\left\{
\begin{array}{ll}
\sigma r  &(r < r_1) 
\\
m_\infty-{Q^2 \over 2 r}  &(r \geq r_1) 
\end{array}
\right. ,
\label{mr}
\eea
where
\bea
m_\infty = \sigma r_1+{Q^2 \over 2 r_1}
\label{minfHH}
\eea
is the ADM mass of the string hedgehog.
We immediately obtain the metric components from \eqref{h} and \eqref{dphidr}, 
\bea
h(r)= e^{-2\phi(r)}=
\left\{
\begin{array}{ll}
{1 \over 1-2\sigma}  &(r < r_1) 
\\
{1 \over \left(1-{r_+ \over r}\right)\left(1-{r_-\over r}\right)} &(r \geq r_1) 
\end{array}
\right. ,
\label{hHH}
\eea
where $r_\pm = m_\infty \pm \sqrt{m_\infty^2 - Q^2}$.
Inside the hedgehog $h$ and $\phi$ are constant, and therefore we have a conical singularity at the origin.

The size $r_1$ is determined by solving the equilibrium equation \eqref{balance}.
Inserting \eqref{rhoHH}, \eqref{prHH} and \eqref{pperpHH} into it we have
\bea
\left[
{\sigma \over 4\pi r^2}-{Q^2 \over 8\pi r^4}
\right]\delta(r-r_1)=0 ,
\eea
and we find
\bea
r_1={Q \over \sqrt{2\sigma}} .
\label{r1Qsigma}
\eea
Here and hereafter $Q$ is understood as the absolute value of the charge.
There is an alternatively and physically more transparent way to determine $r_1$. Indeed one can obtain \eqref{r1Qsigma} by minimizing the ADM mass $m_\infty$ \eqref{minfHH} with respect to $r_1$. We will see that this situation holds for more general cases and use it in later subsections.
By substituting \eqref{r1Qsigma} into \eqref{minfHH}, the ADM mass is given by
\bea
m_\infty
=\sqrt{2\sigma} \, Q .
\eea

\begin{figure}[h]
\begin{center}
\resizebox{6cm}{5cm}{\includegraphics{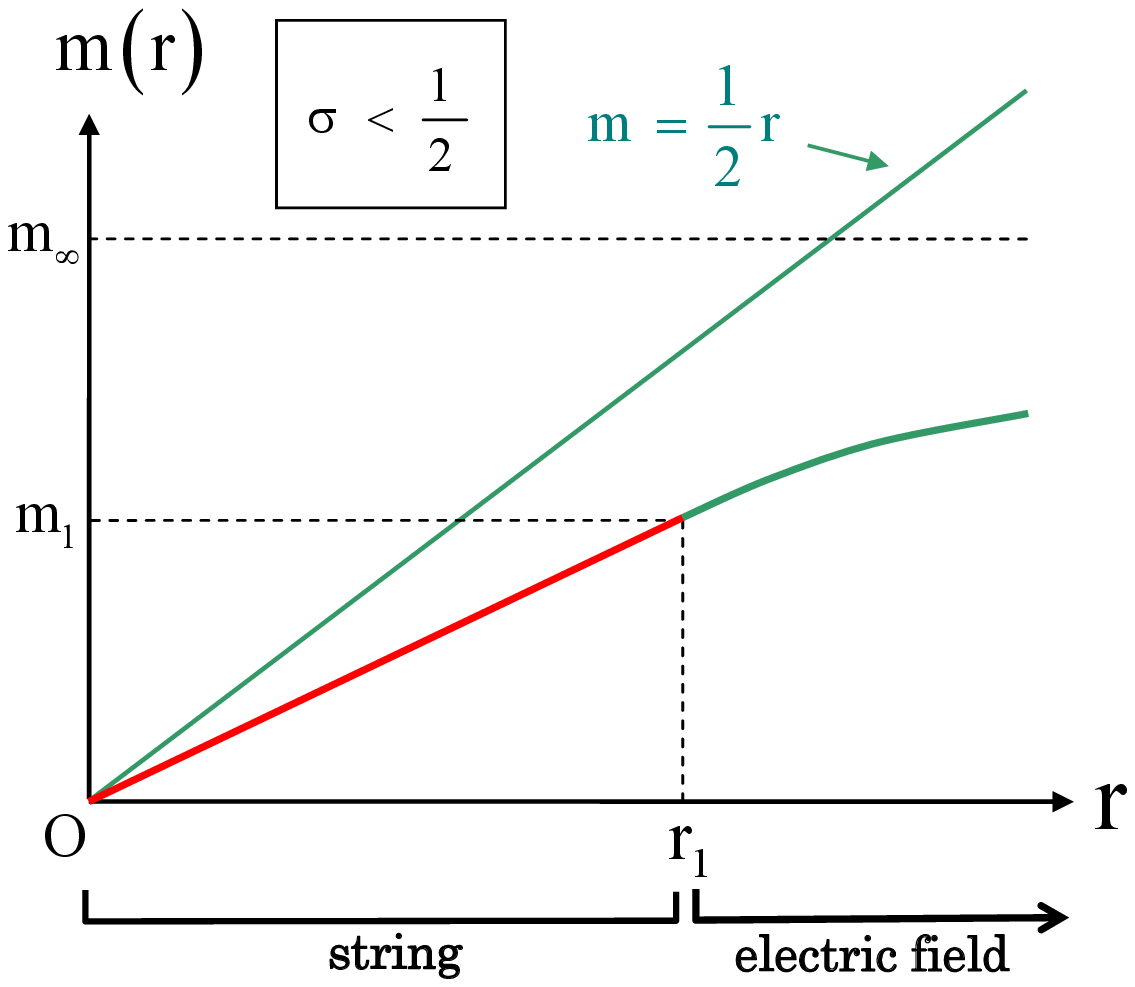}}
\resizebox{6cm}{5cm}{\includegraphics{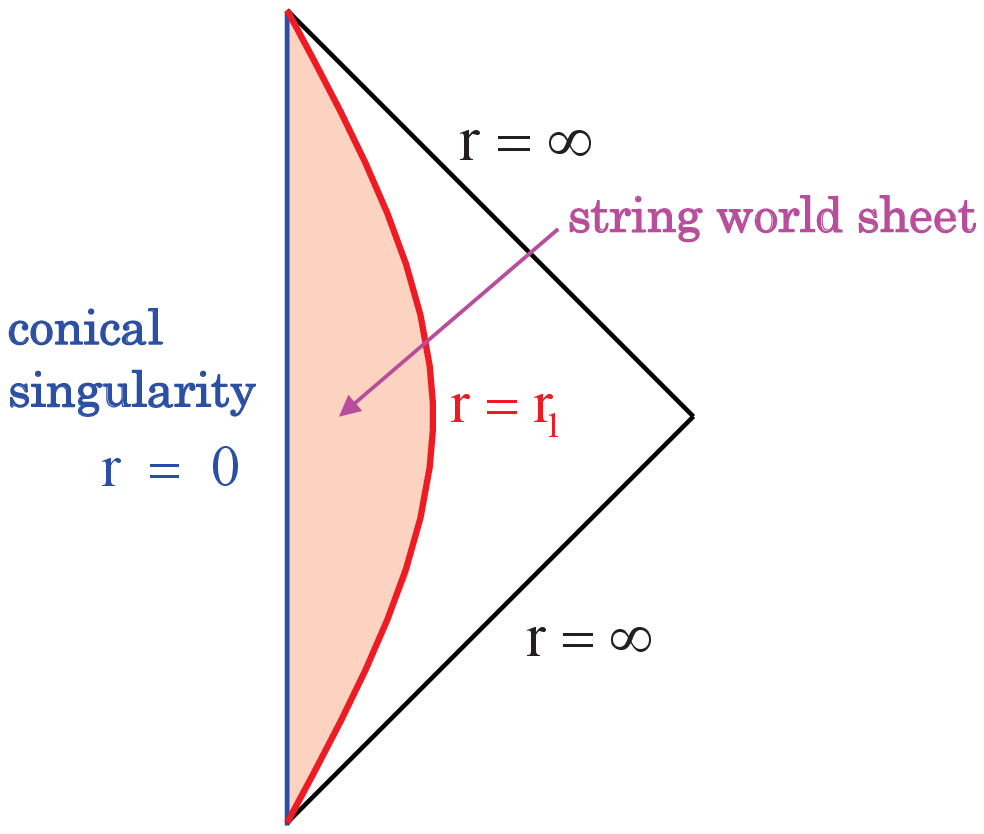}}
\end{center}
\caption{The plot of $m(r)$ of the charged string hedgehog (left) and the Penrose diagram (right) in the case $\sigma<1/2$.}
\label{fig:mVSr_stringOnly1}
\end{figure}

Next we examine the behavior of the metric given by \eqref{mr} and \eqref{hHH}. 
When $\sigma <1/2$, we find $m(r) < r/2$ and thus $h(r)$ is positive everywhere, which means there is no horizon. The metric is non-singular except for the origin. 
In this case we have $m_\infty <Q$, and $r_\pm$ are not real.
The configuration is not like a black hole but like a star (see Figure \ref{fig:mVSr_stringOnly1}).
As $\sigma$ approaches $1/2$, $m_\infty$ becomes close to $Q$, and $h$ becomes infinitely large for $r<r_1$, while the metric in the region $r\geq r_1$ approaches the extremal Reissner-Nordstr\"om solution.
If $\sigma$ is very close to $1/2$, outside observers can hardly distinguish the configuration from the extremal Reissner-Nordstr\"om black hole in the sense that it takes very long time for a particle to get close to the edge where the metric is different from the black hole.

\begin{figure}[h]
\begin{center}
\resizebox{6cm}{5cm}{\includegraphics{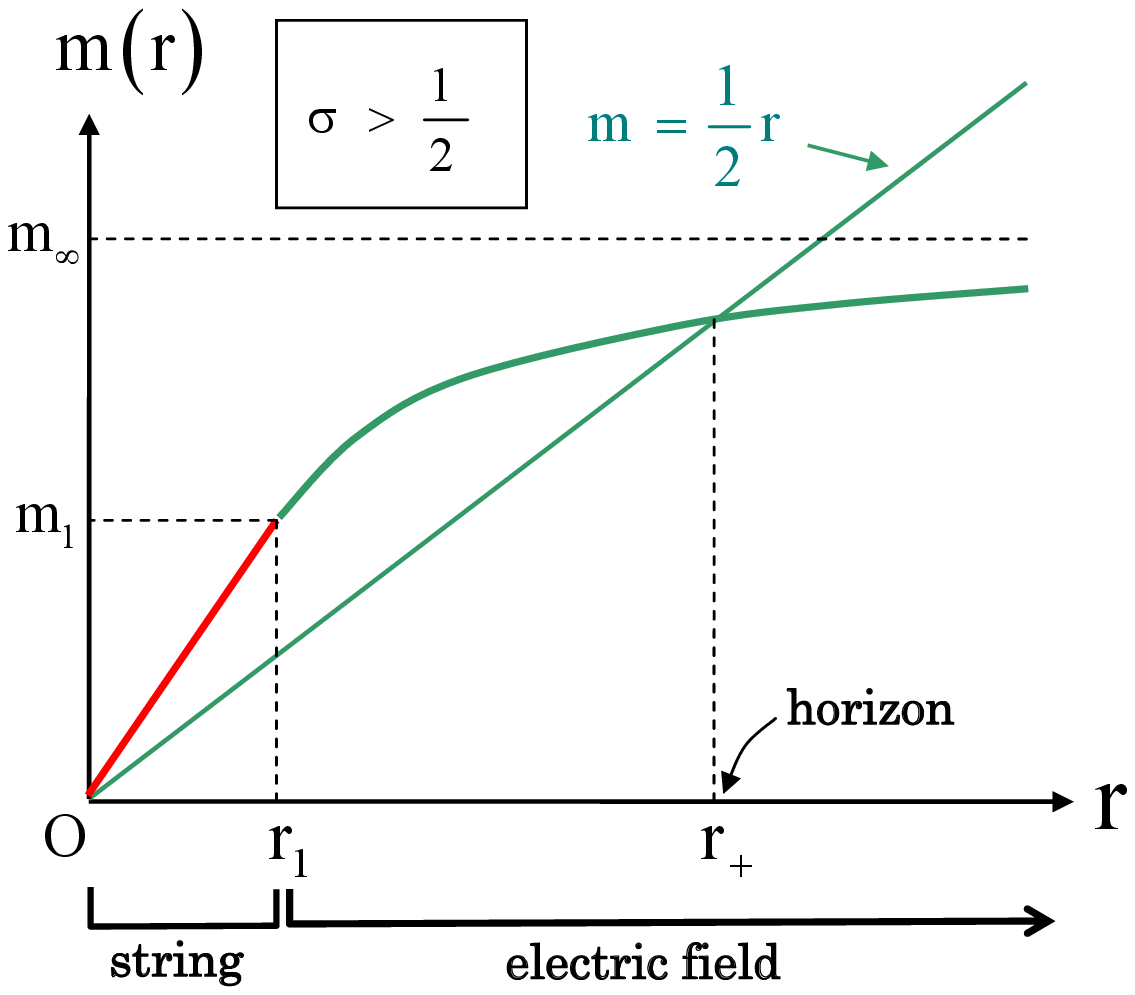}}
\resizebox{6cm}{4.3cm}{\includegraphics{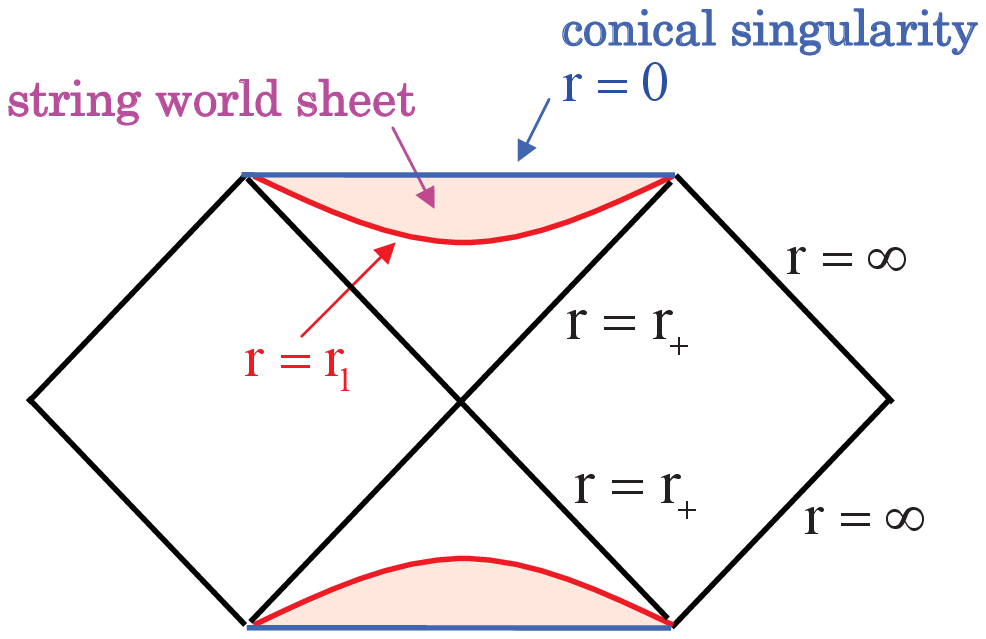}}
\end{center}
\caption{The plot of $m(r)$ of the charged string hedgehog (left) and the Penrose diagram (right) in the case $\sigma>1/2$.
There are two string world sheets, one in the upper triangle and the other in the lower one, in the Penrose diagram, but they are not necessary the same because the two triangle regions are independent. Actually we may replace, for example, the lower triangle by the one that appears in the usual Schwarzschild geometry in that there is the ordinary singularity and no string at the base of the triangle.}
\label{fig:mVSr_stringOnly3}
\end{figure}

On the other hand, when $\sigma \geq 1/2$, we have a horizon.\footnote{Rigorously speaking, we should avoid the coordinate singularity by choosing a new coordinate system like the Kruskal coordinates. However, we continue to use the Schwarzschild coordinates for the sake of simplicity.}
The metric components change their signs at $r=r_+$ as is seen from \eqref{hHH}.
The geometry in the region $r\geq r_1$ coincides with the Reissner-Nordstr\"om black hole solution.
However, $r_-$ has no physical meaning though it is real, because $r_-$ is smaller than $r_1$ and the geometry in the region $r < r_1$ is not described by the Reissner-Nordstr\"om, but a conical geometry.
Thus we conclude that the configuration is a black hole which has only the outer horizon at $r=r_+$ and a conical singularity at the origin.
It has the same causal structure as the Schwarzschild geometry.
Note that the trajectories of the end points of the strings lie on the surface $r=r_1$ which now should be regarded as spacelike, because the time and space directions are interchanged inside the horizon. 
%
What happens is that the strings are produced from the electric field at ``time"  $r=r_1$.
See the Penrose diagram in Figure \ref{fig:mVSr_stringOnly3}.

So far we have considered string hedgehog configurations, in which the density and the pressure diverge at the origin, and the spacetime has a conical singularity. In the following subsections, we show that such singularity can be removed by introducing a membrane that encloses the origin to make its interior flat. 

\subsection{Charged membrane}
Before introducing a membrane to the charged string hedgehog, we consider a system
consisting of a membrane and charges without strings.\footnote{For studies on (un)charged spherical shells in gravitational theory, see for Refs. \citen{Israel:1966,Israel:1967,Bekenstein:1971ej,Chase:1970,Boulware:1973}, and for more recent studies see e.g. Refs. \citen{Guendelman:2008ip,Belinski:2008bn,Bicak:2010zz} and references therein.} 
Let us consider a spherical static membrane with tension $\kappa$ which is charged uniformly with net charge $Q$.
We assume that the membrane is balanced at radius $r_0$ which will be determined below. 
\begin{figure}[h]
\begin{center}
\resizebox{5cm}{5cm}{\includegraphics{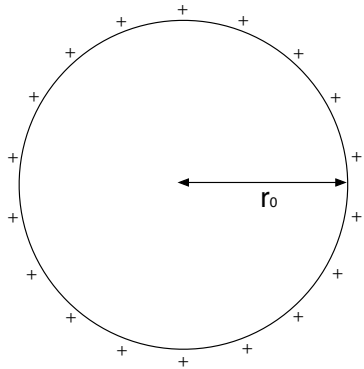}}
\end{center}
\caption{The charged spherical membrane.}
\label{fig:membrane}
\end{figure}
%

%
More precisely, we introduce charged particles that are constrained to move on the membrane, and assume that they couple to the Maxwell field. Then they spread uniformly on the membrane in order to lower the energy. Furthermore, for the sake of simplicity, we set the mass of the particles to zero, so that the energy momentum tensor simply consists of two contributions, one from the Nambu-Goto action of the membrane and the other from the Maxwell field.

The energy-momentum tensor of the membrane is given in Appendix \ref{emtmemb} and that of the Maxwell field in \eqref{MaxwellEMT}. 
Therefore we have
\bea
\rho(r) &=& {\kappa \over \sqrt{h(r_0)}}\delta(r-r_0)+ {Q^2 \over 8 \pi r^4}\theta(r-r_0) ,
\label{dmemb}
\\P_r(r)&=&- {Q^2 \over 8 \pi r^4}\theta(r-r_0) ,
\label{prmemb}
\\P_\perp(r)&=&-{\kappa \over \sqrt{h(r_0)}}\delta(r-r_0) + {Q^2 \over 8 \pi r^4}\theta(r-r_0).
\label{ppmemb}
\eea
In order for the energy-momentum tensor to be real, the condition $h(r_0) > 0$ should be satisfied. 
Also, it is clear that $h(r)$ is not continuous at $r=r_0$ because it is determined through \eqref{h} and \eqref{m} with the density \eqref{dmemb} that contains a delta function.
The ADM mass is in principle obtained by evaluating $m(\infty )$ from \eqref{m} and \eqref{dmemb} as 
\bea
m_{\infty } = \int_0^\infty  dr 4\pi r^2 \left[{\kappa \over \sqrt{h(r_0)}}\delta(r-r_0)
+ {Q^2 \over 8 \pi r^4}\theta(r-r_0)\right] .
\eea
While the second term is easily evaluated, the first term is rather subtle because $h(r)$ has a discontinuity at $r=r_0$.
In order to circumvent it, we start with the equation,
\bea
{d \over dr} m(r) = 4\pi r^2 \rho(r) .
\label{mdiffeq}
\eea
Here $\rho(r)$ is given by \eqref{dmemb}, and contains $h(r)$, which is
expressed in terms of $m(r)$ as \eqref{h}. 
Multiplying $\sqrt{h(r)}$ and integrating over a small region around $r_0$, from $r=r_0-\epsilon$ to $r=r_0+\epsilon$, we obtain
\bea
\int_{r_0-\epsilon}^{r_0+\epsilon} dr {1 \over \sqrt{1- {2 m(r) \over r}}} {d m \over dr} 
=
\int_{r_0-\epsilon}^{r_0+\epsilon} dr
\left[
4\pi \kappa r^2 \delta(r-r_0)+{Q^2 \over 2 r^2} \sqrt{h(r)}\theta(r-r_0)
\right] .
\eea
In the small $\epsilon$ limit, we may safely replace $2m(r)/r$ with $2m(r)/r_0$ in the left hand side, and the integration of the second term in the right hand side vanishes because it is a finite function.
Performing the integration we thus find
\bea
\sqrt{1-{2m(r_0+\epsilon) \over r_0}}-\sqrt{1-{2m(r_0-\epsilon) \over r_0}}
=-4\pi \kappa r_0 .
\label{membranemasseq}
\eea
Because there is no matter inside the membrane, we set $m(r_0-\epsilon)=0$,
and obtain the mass of the membrane,
\bea
m_0=4\pi \kappa r_0^2\left(1- 2\pi \kappa r_0 \right) ,
\label{mpCmemb}
\eea
where $m_0=\lim_{\epsilon \to 0} m(r_0+\epsilon)$.
This equation indicates that the mass of the membrane is reduced by the gravitational binding energy. 
Adding the contribution from the electric field, we obtain the ADM mass of the whole system,
\bea
m_\infty=4\pi \kappa r_0^2\left(1-2\pi \kappa r_0 \right)+{Q^2\over 2 r_0} .
\label{minfCM0}
\eea

\begin{figure}[h]
\begin{center}
\resizebox{8.5cm}{7cm}{\includegraphics{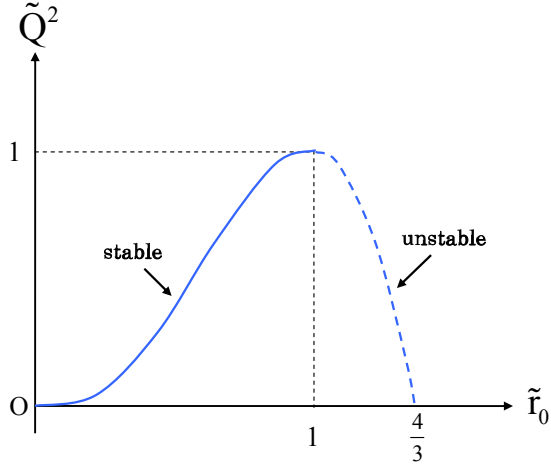}}
\end{center}
\caption{The plot of $\tilde{Q}^2=\tilde{r}_0^3(4-3\tilde{r}_0)$.}
\label{fig:Q2VSr0}
\end{figure}

Now we turn to the problem of finding $r_0$. For this purpose, it is convenient to rescale the variables as
\bea
\tilde{m}_\infty := 4\pi \kappa m_\infty , \quad 
\tilde{r}_0 := 4\pi \kappa r_0 , \quad 
\tilde{Q} := 4\pi \kappa Q , 
\eea
so that \eqref{minfCM0} becomes
\bea
\tilde{m}_\infty=\tilde{r}_0^2\left(1-{\tilde{r}_0 \over 2} \right)+{\tilde{Q}^2\over 2 \tilde{r}_0} .
\label{minfCM}
\eea
Although the radius $r_0$ can be determined by solving the Einstein equation as is done in the previous section, it is more convenient and physically transparent to employ the minimization of the ADM mass with respect to $\tilde{r}_0$. 
We show the equivalence of the two procedures in Appendix \ref{eom}.
As is shown in Figure \ref{fig:minfVSr0}, the function \eqref{minfCM} has two extrema when $\tilde{Q} < 1$, and no extremum when $\tilde{Q} > 1$.
\begin{figure}[h]
\begin{center}
\resizebox{7cm}{6.7cm}{\includegraphics{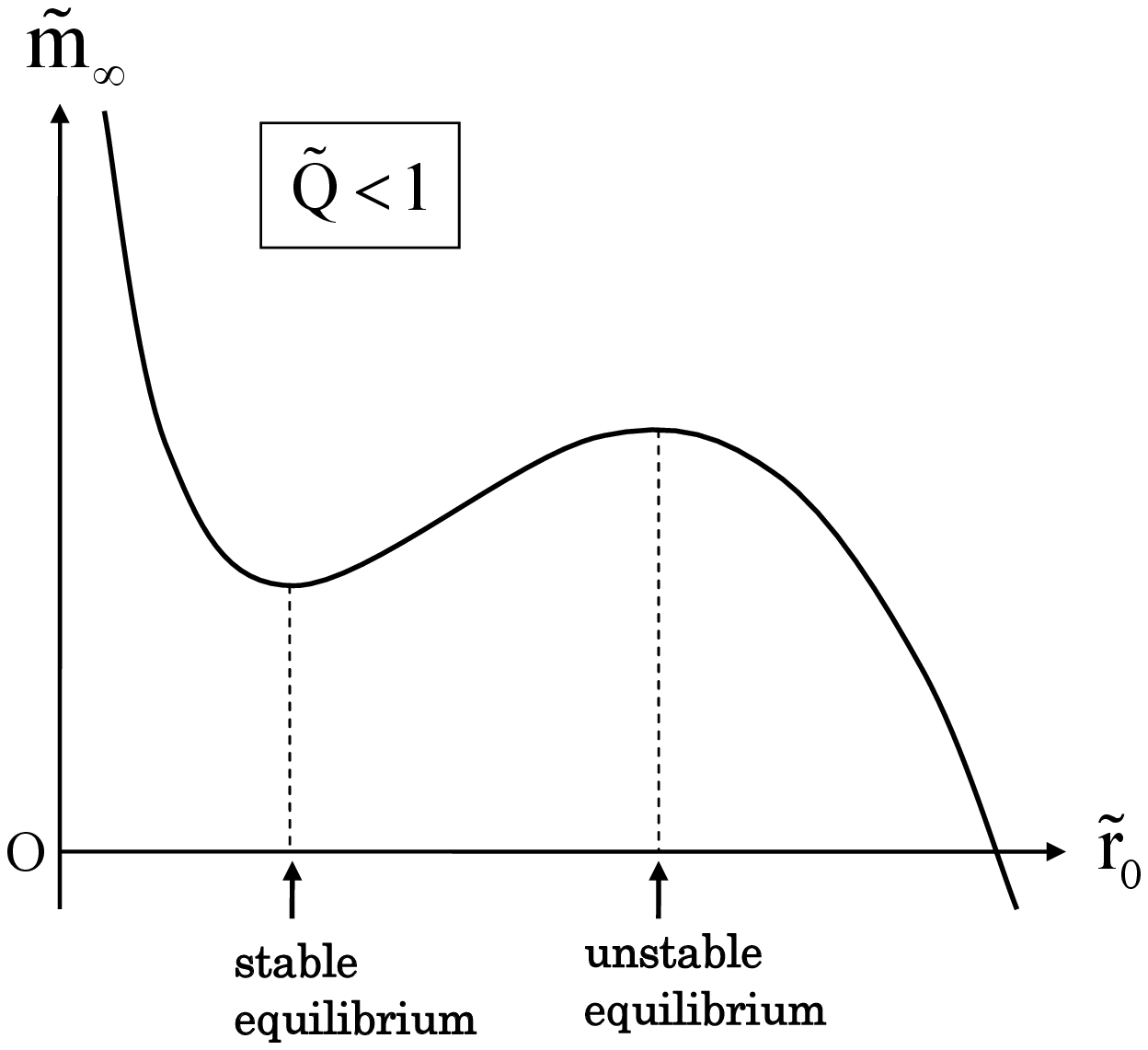}}
\resizebox{6.7cm}{6cm}{\includegraphics{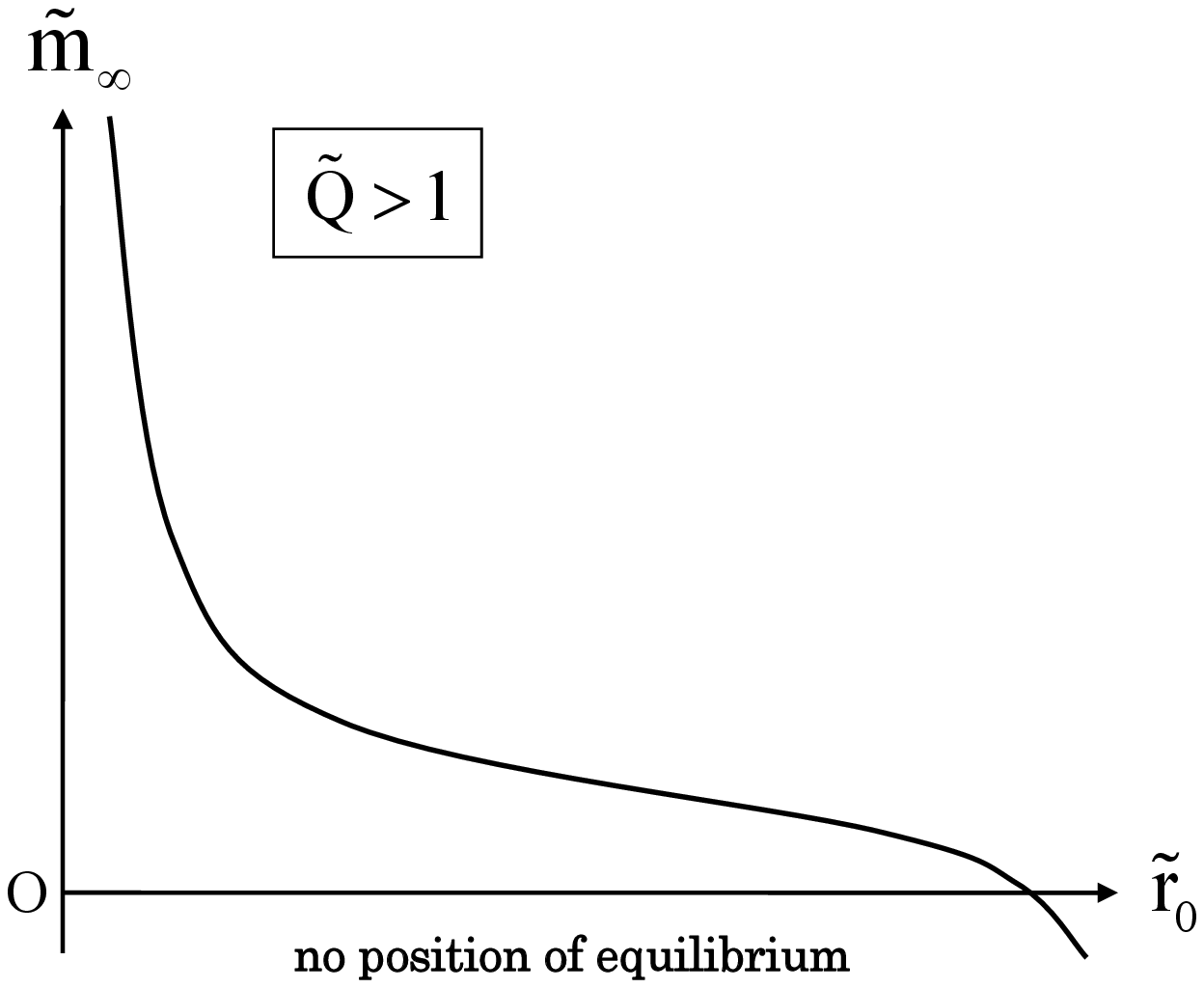}}
\end{center}
\caption{The plots of $\tilde{m}_\infty$ as functions of $\tilde{r}_0$ for $\tilde{Q} <1$ (left) and $\tilde{Q} > 1$ (right).}
\label{fig:minfVSr0}
\end{figure}
The extremal points are determined by $d\tilde{m}_\infty /d\tilde{r}_0=0$, which is expressed as
\bea
\tilde{Q}^2 = \tilde{r}_0^3\left(4-3\tilde{r}_0\right) .
\label{admmvary}
\eea
As is depicted in Figure \ref{fig:Q2VSr0}, we have two positive roots when $\tilde{Q} < 1$, one in the region $0 < \tilde{r}_0 < 1$ and the other in $1 < \tilde{r}_0 <4/3$. 
The former is a local minimum of $\tilde{m}_\infty(\tilde{r}_0)$, and the system is stable, while the latter is a local maximum, and the system is unstable.
Since the solution of the quartic equation \eqref{admmvary} is complicated, we do not try to express various quantities as explicit functions of $\tilde{Q}^2$. Instead, we will see that the equations become simple if we express them in terms of $\tilde{r}_0$.

\begin{figure}[h]
\begin{center}
\resizebox{9cm}{8cm}{\includegraphics{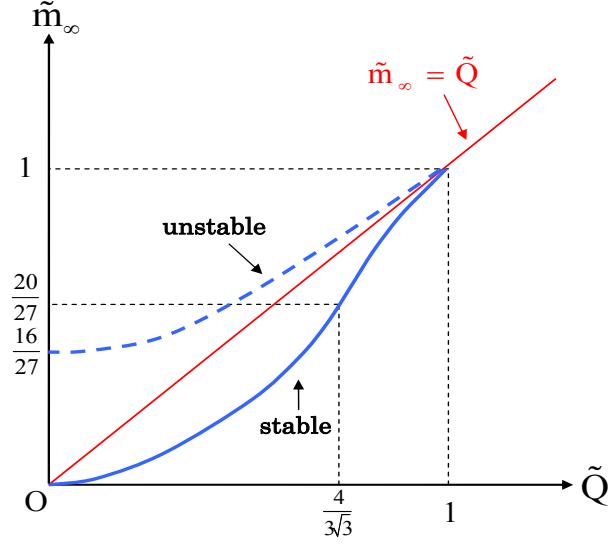}}
\end{center}
\caption{$\tilde{m}_\infty$-$\tilde{Q}$ plot. The curve has a inflection point at $(\tilde{Q}, \tilde{m}_\infty)=(4/(3\sqrt{3}), 20/27)$ where $\tilde{r}_0=2/3$ and is tangent to the slope $\tilde{m}_\infty=\tilde{Q}$ at $\tilde{Q}=1$ where $\tilde{r}_0=1$. 
The slope $\tilde{m}_\infty=\tilde{Q}$ represents the extremal relation.
There is also an unstable branch given by $1 < \tilde{r}_0 \leq 4/3.$}
\label{fig:mQmemb}
\end{figure}

Indeed the ADM mass is obtained by substituting \eqref{admmvary} into \eqref{minfCM} as
\bea
\tilde{m}_\infty = \tilde{r}_0^2(3-2\tilde{r}_0) .
\label{admmass}
\eea
The equations \eqref{admmass} and \eqref{admmvary} can be regarded as a parametric representation of the $\tilde{m}_\infty$-$\tilde{Q}$ curve, which is plotted in Figure \ref{fig:mQmemb}.
The solid and dotted curves correspond to $0 < \tilde{r}_0 < 1$ and 
$1 < \tilde{r}_0 <4/3$, and represent stable and unstable configurations, respectively.
For later use we calculate
\bea
{d \tilde{m}_\infty \over d\tilde{Q}} =\sqrt{4\tilde{r}_0-3\tilde{r}_0^2}, 
\label{dmdQ}
\eea
and 
\bea
{d^2 \tilde{m}_\infty \over d\tilde{Q}^2} ={2-3\tilde{r}_0 \over 6\tilde{r}_0(1-\tilde{r}_0)},
\eea
which shows that the $\tilde{m}_\infty$-$\tilde{Q}$ curve has an inflection point at $(\tilde{Q}, \tilde{m}_\infty)=(4/(3\sqrt{3}), 20/27)$ where $\tilde{r}_0=2/3$. 
Note that it is also an inflection point of the mass of the membrane \eqref{mpCmemb}.
We can also show $\tilde{m}_\infty < \tilde{Q}$ for  $0<\tilde{r}_0<1$, and
$\tilde{m}_\infty > \tilde{Q}$ for $1<\tilde{r}_0<4/3$.
In the following discussions, we will consider only stable configurations, $0<\tilde{r}_0\leq 1$.

\begin{figure}[h]
\begin{center}
\resizebox{8cm}{7cm}{\includegraphics{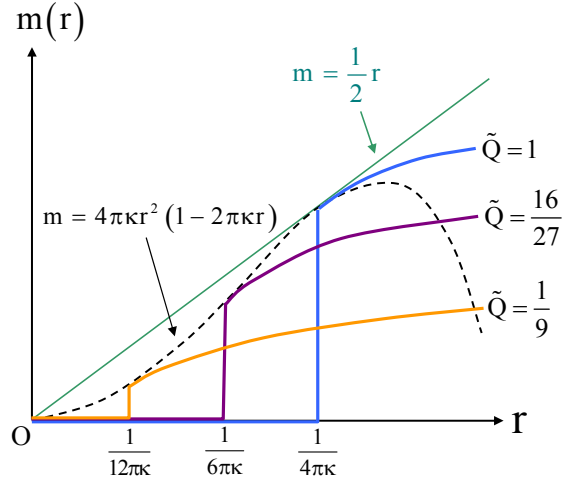}}
\end{center}
\caption{The plot of the functions $m(r)$ of the charged membrane for $\tilde{Q}=1~ (\tilde{r}_0=1)$, $\tilde{Q}=16/27~(\tilde{r}_0=2/3)$ and $\tilde{Q}=1/9~(\tilde{r}_0=1/3)$.
These curves are tangent to the curve of the membrane mass $m=4\pi \kappa r^2(1-2\pi \kappa r)$.}
\label{fig:mVSr_braneOnly}
\end{figure}

The metric is also determined by a simple calculation. 
Because there is nothing in the region $r < r_0$, and only the electric field in 
$r > r_0$, we have 
\bea
m(r)=\left(m_\infty-{Q^2 \over 2 r}\right)\theta(r-r_0) .
\label{m(r)memb}
\eea
Then from \eqref{h} we have
\bea
h(r)
=
\left\{
\begin{array}{ll}
1  &(r < r_0) 
\\
{1\over 1-{2m_\infty \over r} + {Q^2 \over r^2}}  &(r \geq r_0) 
\end{array}
\right. ,
\label{h(r)memb}
\eea
and from \eqref{dphidr}, \eqref{prmemb}, \eqref{m(r)memb} and \eqref{h(r)memb} we obtain
\bea
e^{2\phi(r)}
=
\left\{
\begin{array}{ll}
1-{2m_\infty \over r_0} + {Q^2 \over r_0^2} &(r < r_0) 
\\
1-{2m_\infty \over r} + {Q^2 \over r^2} &(r \geq r_0) 
\end{array}
\right. .
\eea
Note that $\phi(r)$ is continuous at $r=r_0$, while $h(r)$ is not \cite{Jacobson:2007tj}.
The function $m(r)$ for various charges (radius) are plotted in Figure \ref{fig:mVSr_braneOnly}.
To summarize, if we consider stable configurations, the charged membrane has no horizons except for the extremal limit $\tilde{Q}=1$. Furthermore there is no singularity even in the extremal limit.

Before closing this subsection, it is worth mentioning that the system considered here can be regarded as a generalization of the Poincar\'e-Schwinger (Abraham-Lorentz) model of electron in which gravity is included (though spin is not included).
However, we find that the effect of gravity does not help, because the radius of the membrane is always larger than the classical electron radius, as is obvious from \eqref{minfCM},
\bea
r_0={4\pi \kappa r_0^3 \over m_\infty}\left(1-2\pi \kappa r_0 \right) + {Q^2 \over 2 m_\infty} 
\geq {Q^2 \over 2 m_\infty} = r_{classical} .
\eea
Note that the membrane mass $4\pi \kappa r_0^2 \left(1-2\pi \kappa r_0 \right)$ is always positive, because $4\pi \kappa r_0<4/3$ even if we allow unstable configurations. The gravitational interaction decreases the mass but does not make it negative.

\subsection{String-membrane hedgehog}

Now we use the membrane solution just studied in the last subsection to remove the conical singularity in the charged hedgehog configuration we found in subsection \ref{String hedgehog}.
In this way, we shall obtain a configuration that has horizons without any singularity at the origin.

\begin{figure}[h]
\begin{center}
\resizebox{5.5cm}{6cm}{\includegraphics{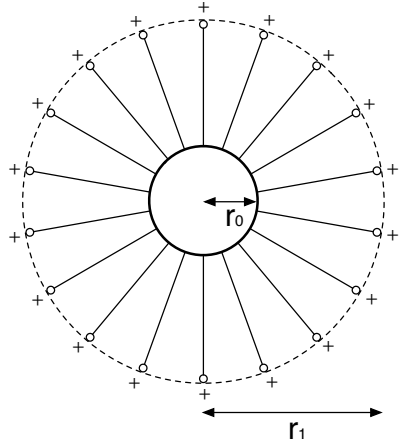}}
\end{center}
\caption{The charged hedgehog with membrane inside. Two dimensional slice.}
\label{fig:hedgehog-membrane}
\end{figure}

We insert a neutral spherical membrane around the origin, which is connected to the strings in a charged hedgehog \cite{Guendelman:1991qb}.
The membrane tends to shrink to the origin while the strings pull it radially so that the membrane is suspended.
In addition, the strings are balanced with the Coulomb force at the other end points. See Figure \ref{fig:hedgehog-membrane}.

The energy density and the pressure consist of three contributions, one from the membrane, one from the strings, and one from the electric field,
\bea
\rho(r) &=& {\kappa \over \sqrt{h(r_0)}}\delta(r-r_0)
+{\sigma \over 4 \pi r^2} \theta(r-r_0)\theta(r_1-r)
+{Q^2 \over 8 \pi r^4}\theta(r-r_1) ,
\label{rhoStringMembrane}
\\P_r(r)&=&-{\sigma \over 4 \pi r^2} \theta(r-r_0)\theta(r_1-r)
-{Q^2 \over 8 \pi r^4}\theta(r-r_1) ,
\label{PrStringMembrane}
\\P_\perp(r)&=&-{\kappa \over \sqrt{h(r_0)}}\delta(r-r_0)+{Q^2 \over 8 \pi r^4}\theta(r-r_1) ,
\eea
where $r_0$ and $r_1$ are the positions of the inner and outer end points of the strings, respectively.
The ADM mass also consists of three pieces,
\bea
m_\infty=4\pi \kappa r_0^2\left(1-2\pi \kappa r_0 \right)+\sigma(r_1-r_0)+{Q^2\over 2 r_1} ,
\label{ADMorigi}
\eea
where the first term is the membrane mass \eqref{mpCmemb}, the second term is the mass of the strings, and the third term is the contribution from the electric field.

As in the case of the charged membrane we introduce rescaled variables,
\bea
\tilde{m}_\infty = 4\pi \kappa m_\infty , \quad 
\tilde{r}_0 = 4\pi \kappa r_0 , \quad 
\tilde{r}_1 = 4\pi \kappa r_1 , \quad 
\tilde{Q} = 4\pi \kappa Q ,
\label{redefStringMembrane}
\eea
in which \eqref{ADMorigi} becomes
\bea
\tilde{m}_\infty=\tilde{r}_0^2\left(1-{\tilde{r}_0\over 2}\right)+\sigma(\tilde{r}_1-\tilde{r}_0)+{\tilde{Q}^2\over 2 \tilde{r}_1} .
\label{ADMredef}
\eea
By taking variations with respect to $\tilde{r}_0$ and $\tilde{r}_1$, we find that $\tilde{m}_\infty$ is minimized at
\bea
\tilde{r}_1={\tilde{Q} \over \sqrt{2 \sigma}} ,
\label{r1}
\eea
and
\bea
\tilde{r}_0={2 \over 3} \left(1 - \sqrt{1-{3 \over 2}\sigma}\right) .
\label{r0}
\eea
In \eqref{r0} we have selected the solution of
\bea
\sigma={\tilde{r}_0 \over 2} \left(4-3\tilde{r}_0\right) ,
\label{sigma-r0}
\eea
that gives the local minimum of $\tilde{m}_\infty$, so that the system is stable.
We also have a necessary condition for the existence of a minimum,
\bea
\sigma  \leq {2\over 3} .
\label{sigma_lt_2/3}
\eea
This inequality means that if the string tension exceeds $2/3$, the membrane will be torn by the strings no matter how large the membrane tension is.
From \eqref{r0} we find that the radius of the membrane is bounded as
\bea
\tilde{r}_0 \leq {2 \over 3} .
\eea 
Furthermore, from the inequality $\tilde{r}_1 \geq \tilde{r}_0$, we have 
\bea
\tilde{Q} \geq \sqrt{2 \sigma}\tilde{r}_0.
\label{Qbound}
\eea
 
\begin{figure}[h]
\begin{center}
\resizebox{9cm}{7cm}{\includegraphics{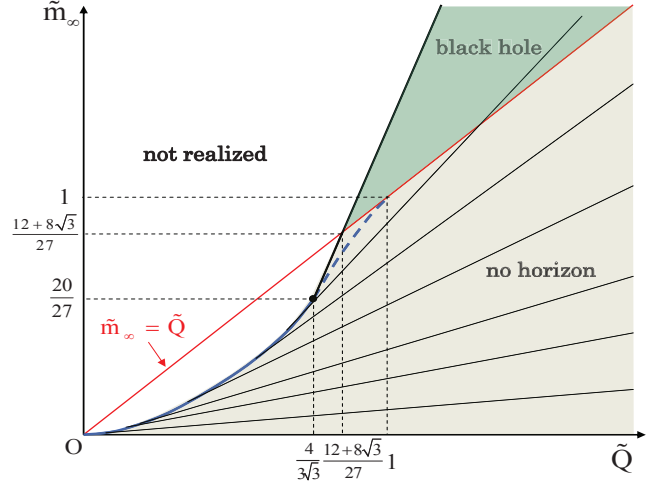}}
\end{center}
\caption{$\tilde{m}_\infty$-$\tilde{Q}$ graph of the hedgehog-membrane. The shaded are the regions that can be realized by the hedgehog-membrane. Horizons appear in the region above the line $\tilde{m}_\infty = \tilde{Q}$.}
\label{fig:MQ}
\end{figure}

Substituting \eqref{r1} and \eqref{sigma-r0} back into \eqref{ADMredef}, we find that the ADM mass can be written as
\bea
\tilde{m}_\infty
=-\tilde{r}_0^2(1-\tilde{r}_0)+\sqrt{2 \sigma} \tilde{Q} .
\label{minf}
\eea
This equation has a simple and interesting physical interpretation, if we regard it as expressing the allowed region in the $\tilde{m}_\infty$-$\tilde{Q}$ plane.
For a given $\tilde{r}_0$ (or $\sigma$ by \eqref{sigma-r0}), because of the restriction \eqref{Qbound}, the equation \eqref{minf} represents a half line that starts from the point $(\tilde{Q},\tilde{m}_\infty)=(\sqrt{2\sigma}\tilde{r}_0, 3\tilde{r}_0^2-2\tilde{r}_0^3)$. 
This expression coincides with the parametric representation of the curve of the charged membrane solution in the previous section, namely \eqref{admmvary} and \eqref{admmass}.
Furthermore, the half line \eqref{minf} is tangent to the curve as is seen from \eqref{dmdQ}. 
Therefore, as $\tilde{r}_0$ varies from $0$ to $2/3$, we have a set of half lines whose envelope is the membrane curve as is depicted in Figure \ref{fig:MQ}. 
Recall that the point at $\tilde{r}_0=2/3$ is the inflection point of the curve.
These are naturally understood, if one recognizes that the charged membrane can be obtained from the charged hedgehog-membrane by setting the string tension to a critical value so that the length of the strings becomes zero.
However the charged membrane in the parameter region $2/3 <\tilde{r}_0 <1$ can not be obtained by such limit (the dotted line in Figure \ref{fig:MQ}).
As we will see below, the region above the line $\tilde{m}_\infty=\tilde{Q}$ corresponds to the solutions which should be regarded as black holes. We would like to mention that  we can recover the string hedgehog solutions, if we take the large membrane tension limit. However, it makes sense only under the condition \eqref{sigma_lt_2/3}.

Now it is straightforward to calculate $h(r)$ and $\phi(r)$.
From \eqref{m} and \eqref{rhoStringMembrane}, we have 
\bea
m(r)
=
\left\{
\begin{array}{ll}
0   &(r < r_0) 
\\
m_0 + \sigma (r-r_0)  &(r_0 \leq r <  r_1) 
\\
m_\infty-{Q^2 \over 2 r}  &(r \geq r_1) 
\label{m(r)StringMembrane}
\end{array}
\right. ,
\eea
where $m_0$ and $m_\infty$ ($\tilde{m}_\infty$) are given by \eqref{mpCmemb} and \eqref{minf}, respectively.
Then from \eqref{h}, \eqref{dphidr} and \eqref{PrStringMembrane} it follows that 
\bea
h(r)
=
\left\{
\begin{array}{ll}
1  &(r < r_0) 
\\
{1 \over 1- {2m_0 \over r}-2\sigma \left(1-{r_0\over r}\right)}  &(r_0 \leq r <  r_1) 
\\
{1 \over 1-{2 m_\infty \over r} + {Q^2 \over r^2}}  &(r \geq r_1) 
\end{array}
\right. ,
\eea
and 
\bea
e^{2 \phi(r)}
=
\left\{
\begin{array}{ll}
1- {2m_0 \over r_0}  &(r < r_0) 
\\
1- {2m_0 \over r}-2\sigma \left(1-{r_0\over r}\right)  &(r_0 \leq r <  r_1) 
\\
1-{2 m_\infty \over r} + {Q^2 \over r^2}  &(r \geq r_1) 
\end{array}
\right. .
\eea

If $\sigma<1/2$, it is clear from \eqref{m(r)StringMembrane} that $m(r)<r/2$ everywhere, and  there is no horizon. On the other hand, if $\sigma>1/2$, the function $m(r)$ may exceed the line $r/2$ and horizons may appear depending on the values of $\kappa $ and $Q$.
As we will see below, we can make non-extremal black holes as well as extremal ones.
For a fixed $\sigma$, the solutions are characterized by the rescaled charge $\tilde{Q}$,
and in fact we have the following three cases.

\begin{figure}[h]
\begin{center}
\resizebox{7cm}{7cm}{\includegraphics{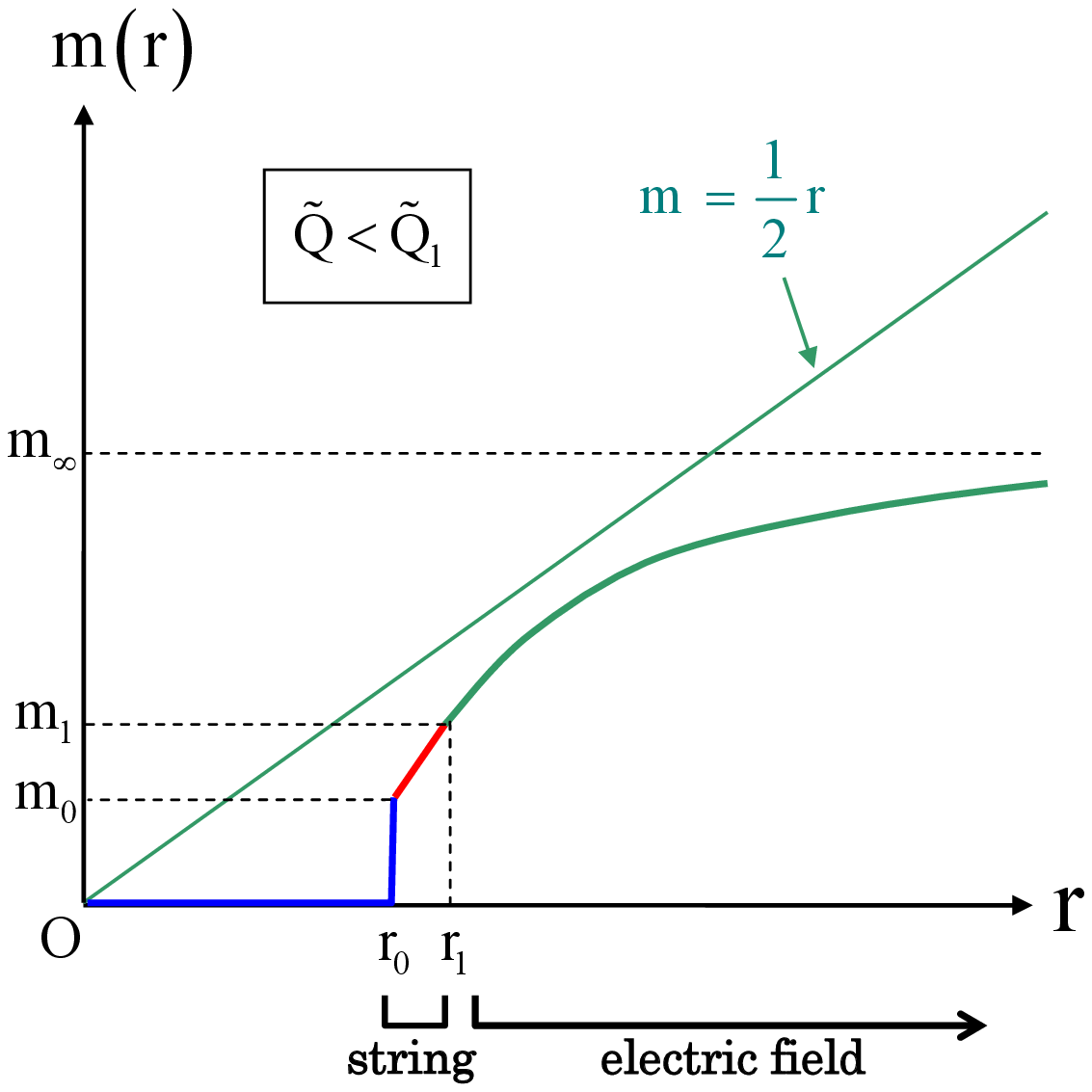}}
\resizebox{6.5cm}{6cm}{\includegraphics{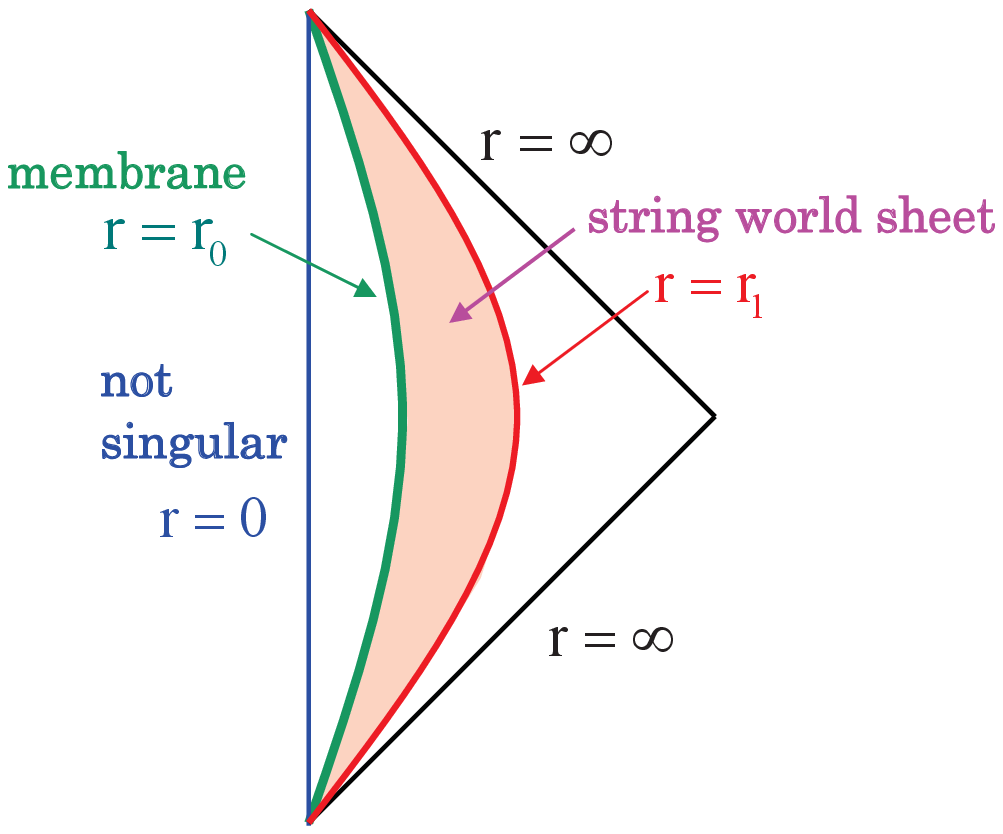}}
\end{center}
\caption{The plot of $m(r)$ of the charged string-membrane hedgehog with a fixed $\sigma$, and $\tilde{Q}<\tilde{Q}_1$ (left), and the corresponding Penrose diagram (right).}
\label{fig:mVSr_string-brane1}
\end{figure}

Firstly, if $\tilde{Q}$ is so small that 
\bea
\tilde{Q} < \tilde{Q}_1,
\eea
where $\tilde{Q}_1={\tilde{r}_0^2(1-\tilde{r}_0) \over \sqrt{2\sigma}-1}$,
$m(r)$ does not exceed $r/2$, and we find no horizon (see Figure \ref{fig:mVSr_string-brane1}).
In this case the configuration is seen as a charged star. There is no singularity at the origin, unlike the case of $\tilde{Q} > \tilde{m}_\infty$ in the Reissner-Nordstr\"om solution.

\begin{figure}[h]
\begin{center}
\resizebox{6.5cm}{6cm}{\includegraphics{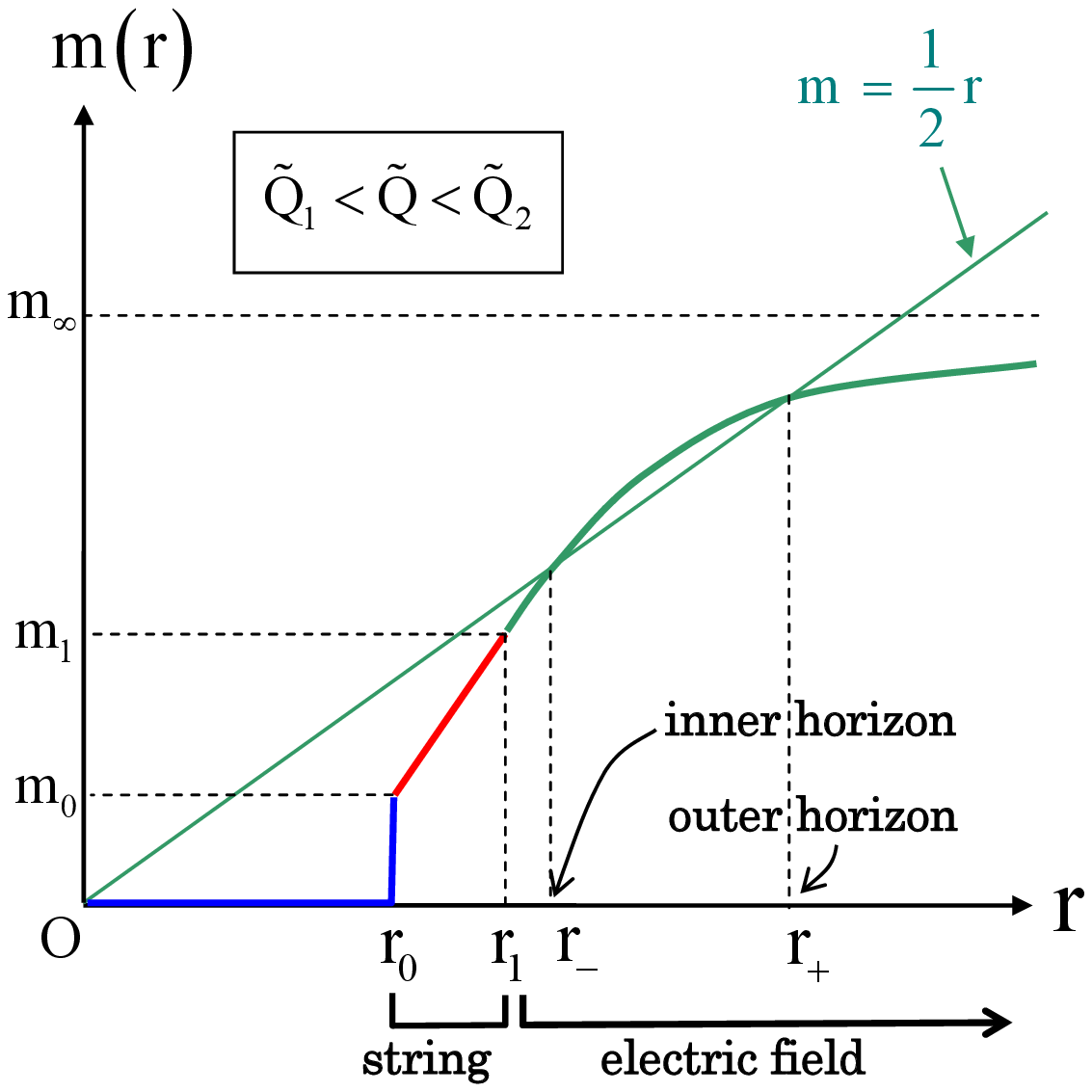}}
\resizebox{6cm}{6cm}{\includegraphics{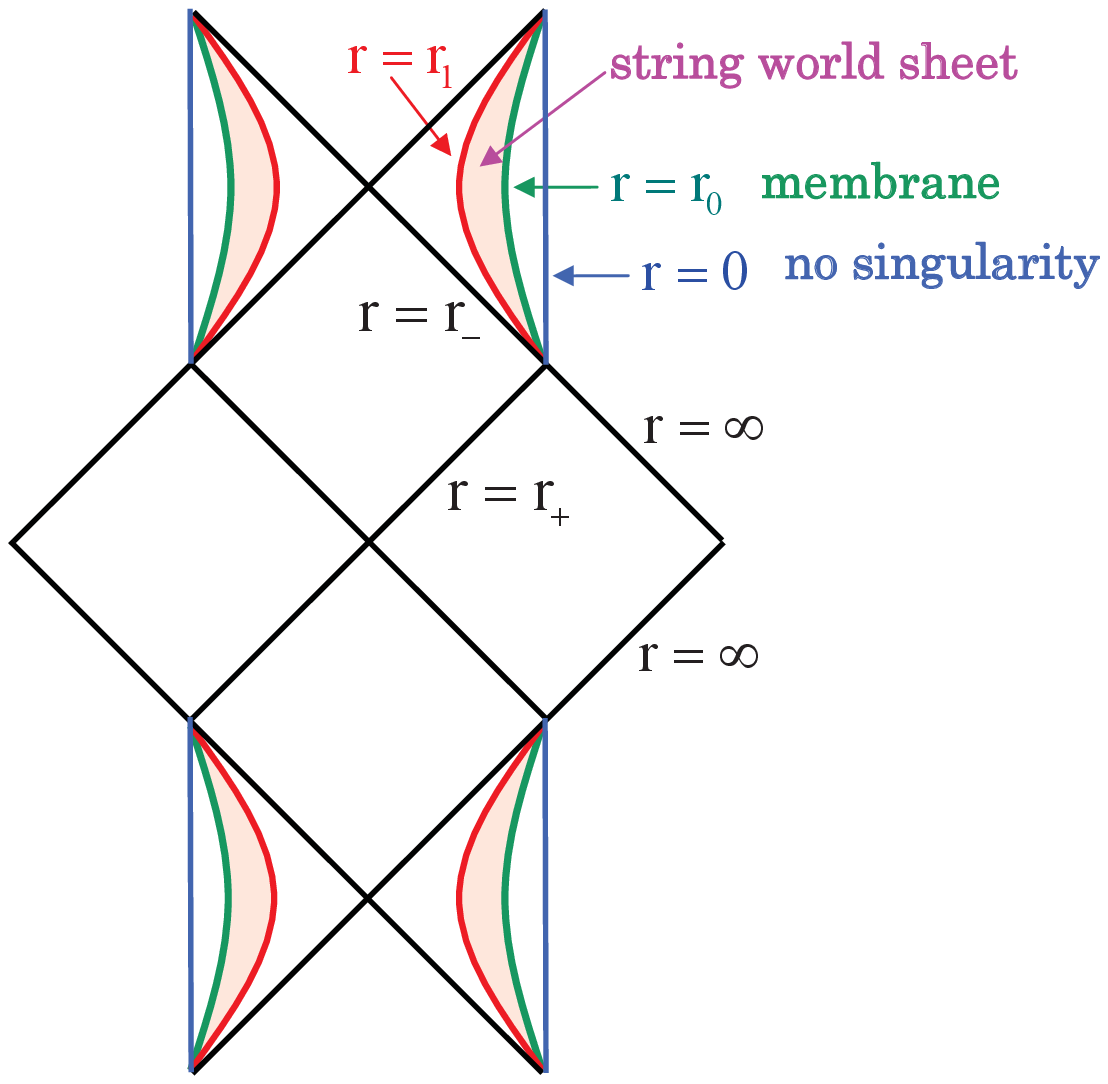}}
\end{center}
\caption{The plot of $m(r)$ of the charged string-membrane hedgehog with a fixed $\sigma$, and $\tilde{Q}_1 < \tilde{Q} < \tilde{Q}_2$ (left), and the corresponding Penrose diagram (right).}
\label{fig:mVSr_string-brane3}
\end{figure}

Secondly, if $\tilde{Q}$ is in the region,
\bea
\tilde{Q}_1  < \tilde{Q} < \tilde{Q}_2,
\label{Qineq}
\eea
where $\tilde{Q}_2={2\tilde{r}_0^2 \sqrt{2\sigma}\over 3\tilde{r}_0-1}$,
the Coulomb potential part of $m(r)$ exceeds $r/2$,
and we have two horizons at $r_+$ and $r_-$ (see Figure \ref{fig:mVSr_string-brane3}).
In this case,  $r_+$ and $r_-$ are the same as those of the Reissner-Nordstr\"om solution.
When $\tilde{Q}=\tilde{Q}_1$, $\tilde{r}_+$ and $\tilde{r}_-$ merge, and we have an extremal black hole.
When $\tilde{Q}=\tilde{Q}_2$,  the outer end points of strings reach the inner horizon, $\tilde{r}_1=\tilde{r}_-$.

%

\begin{figure}[h]
\begin{center}
\resizebox{6.5cm}{6cm}{\includegraphics{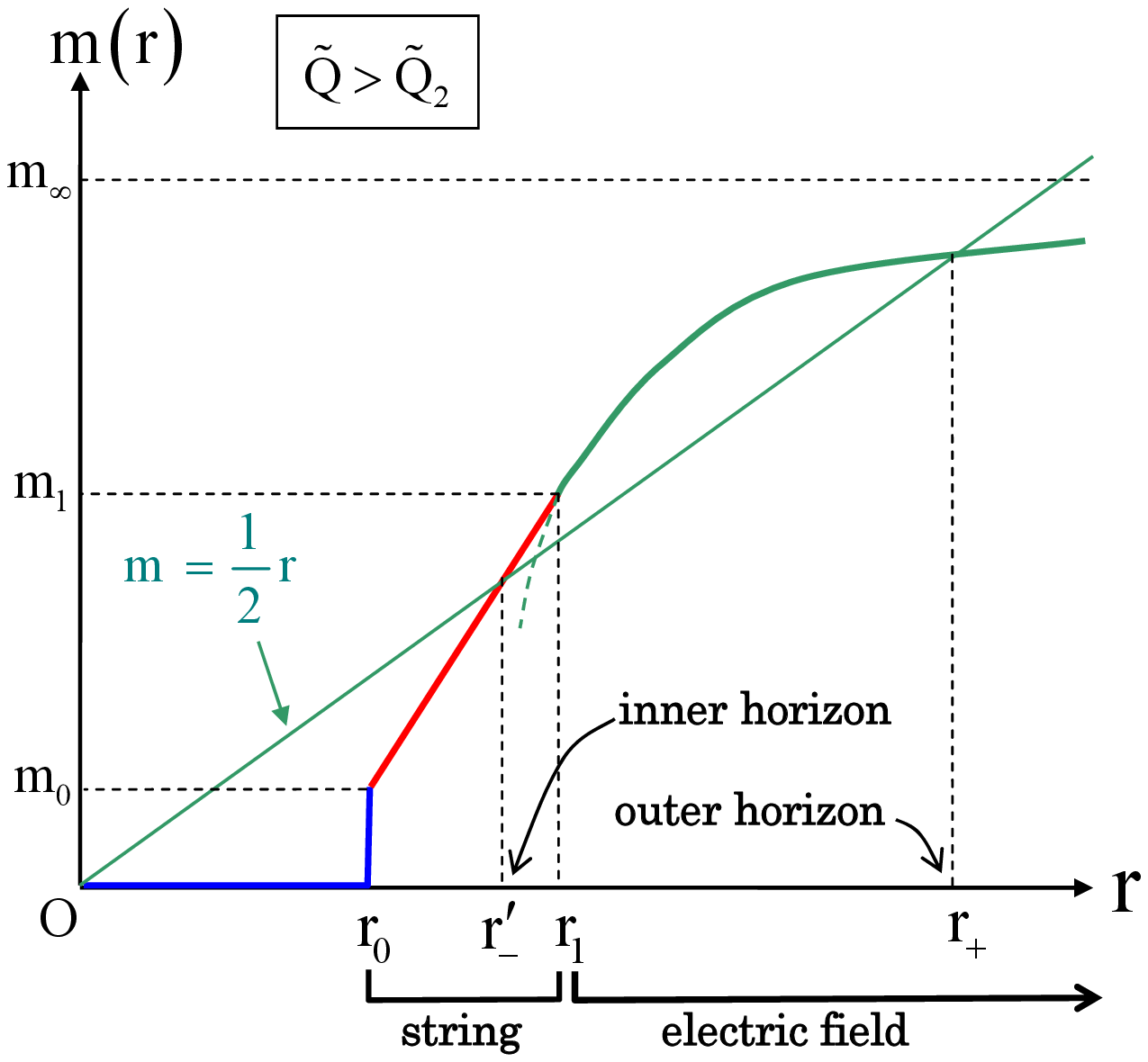}}
\resizebox{6.5cm}{6cm}{\includegraphics{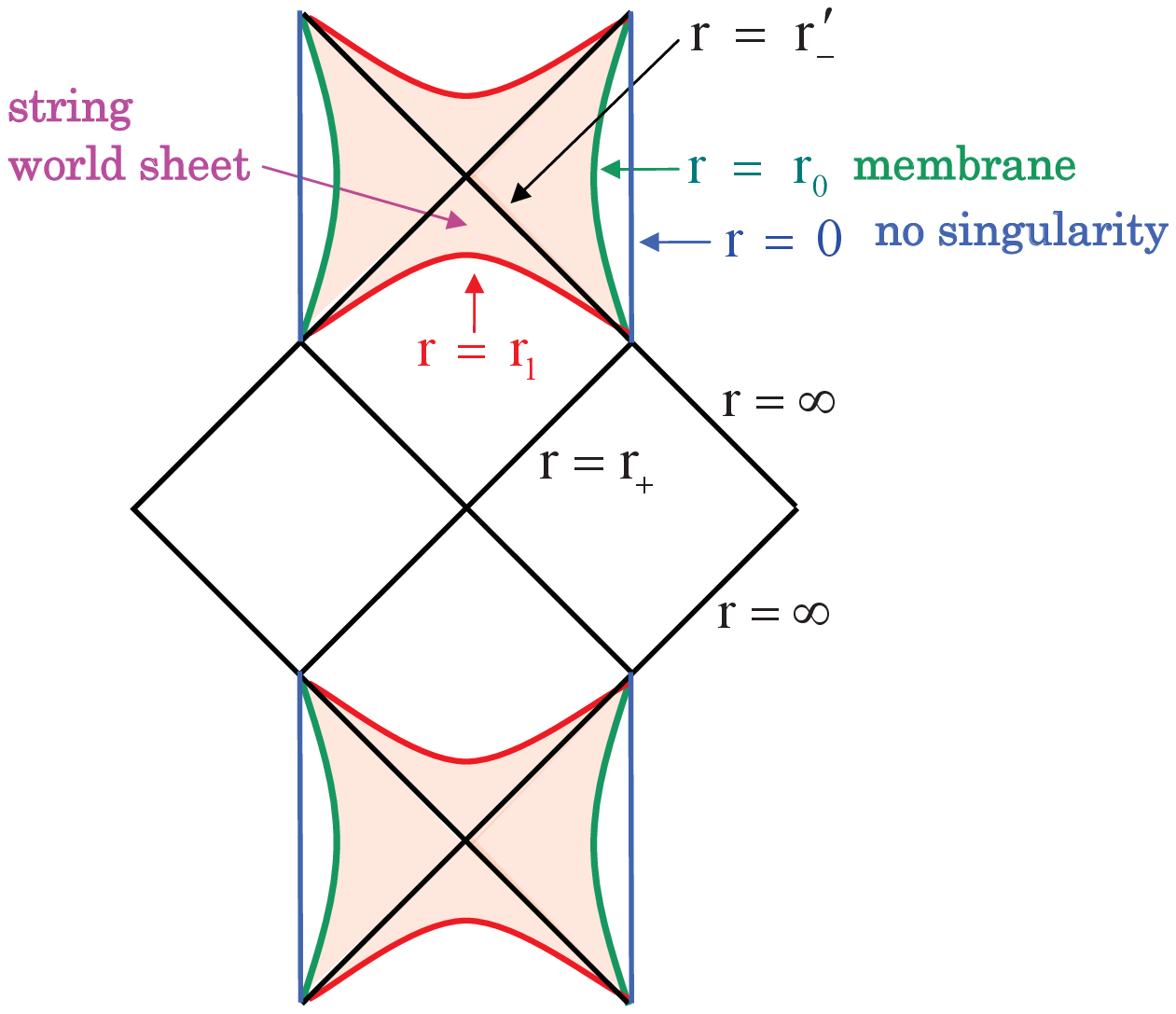}}
\end{center}
\caption{The plot of $m(r)$ of the charged string-membrane hedgehog with a fixed $\sigma$, and $\tilde{Q}_2 < \tilde{Q}$ (left), and the corresponding Penrose diagram (right).}
\label{fig:mVSr_string-brane5}
\end{figure}

Lastly, if $\tilde{Q}$ is so large that
\bea
\tilde{Q}_2  < \tilde{Q} ,
\eea
we have a new type of solutions (see Figure \ref{fig:mVSr_string-brane5}).
Although the outer horizon is still the same as that of the Reissner-Nordstr\"om solution $r_+$,
the inner horizon is now given by the intersection $r_-'$ of the string part of $m(r)$ and $r/2$.
By a simple calculation we have
\bea
r'_-={8\pi \kappa r_0^2(1-4\pi \kappa r_0) \over 2\sigma -1} < r_-.
\eea

A comment is in order here.
Although we have found horizons that are not accompanied with singularities, this is not a contradiction because membranes do not satisfy the strong energy condition on which the singularity theorem is based \cite{Hawking:1973uf}. Actually, the condition can be written as \cite{Wald:1984rg}
\bea
\rho+P_r+2P_\perp \geq 0, \quad {\mbox{and}} \quad \rho+P_i \geq 0 ~ (i=r, \perp),
\label{strongene}
\eea
and as is easily seen from \eqref{dmemb}, \eqref{prmemb} and \eqref{ppmemb}, the first inequality is not satisfied.

\section{Discussion and summary}
Although the configurations we considered in section 2 are static (zero temperature), once we take into account fluctuations around the solutions a lot of degrees of freedom will emerge. Furthermore, a large redshift inside the configurations makes such the degrees of freedom quite huge which might be a source of the entropy of the corresponding black holes. 
In this discussion section we consider, instead of the thermal fluctuation around the solutions, the simplest model of thermal motion to capture the essential point of the problem. We note that the thermal motion can make the strings stretched without charges and the system may have a finite size.


We consider a system consisting of $N$ open strings. This time we assume both ends of the strings are fixed to a point $O$, and we do not introduce any charge. Since the ground state of such system has zero size classically, we consider finite temperature states. 

A long string with finite temperature is well approximated by a random walk. Because the both end points are fixed to the origin, the averaged mass density $\rho(r) $ of the string is given by the square of Green's function $G(r)$,
\bea
\rho(r)=G(r)^2 .
\label{rhoFiniteTemp}
\eea
Here, $G(r)$ is determined by the diffusion equation \cite{Horowitz:1997jc},
\bea
{e^{-\phi(r)} \over r^2 \sqrt{h(r)}} \p_r \left({r^2 e^{\phi(r)} \over \sqrt{h(r)}}  \p_r G(r)\right)
=\left(e^{2\phi(r)} \beta^2 -\beta_H^2\right) G(r) ,
\label{diffusionEq}
\eea
where $\beta$ is the inverse temperature observed at infinity, and $\beta_H$ is the inverse of the Hagedorn temperature that is of order of the string scale.
If $h(r)$ and $\phi(r)$ are constant as in the case of the charged string hedgehog,  \eqref{diffusionEq}
is easily solved as 
\bea
G(r) \propto {e^{-\xi \sqrt{h_0} r} \over r} .
\eea
Here we have assumed $h(r)=h_0$ and $\phi(r)=\phi_0$, and $\xi=e^{2\phi_0} \beta^2 -\beta_H^2$ is a constant.
If the local temperature inside the configuration is at the Hagedorn temperature namely $e^{2\phi_0} \beta^2 -\beta_H^2 =0$, the density \eqref{rhoFiniteTemp} becomes
\bea
\rho \propto {1 \over r^2} .
\eea
If we have $N$ strings, we should multiply it by $N$, but at any rate $\rho$ is proportional to $1/r^2$.
Then the $h(r)$ obtained from \eqref{m} is a constant, which verifies the self consistency of the initial assumption.

In general, the entropy of a thermal string is proportional to its length. 
Therefore, the entropy in our case is given by the proper mass, which is equal to the total intrinsic length of the strings,
\bea
S \simeq
l_s \int_{0}^{r_0} dr 4\pi r^2 \sqrt{h_0} \rho(r) 
= l_s\sqrt{h_0} M ,
\label{stringyEntropy}
\eea
where $l_s$ is the string length scale and $M$ is the ADM mass,
\bea
M =
\int_{0}^{r_0} dr 4\pi r^2 \rho(r) .
\eea
Here we have assumed that $\rho (r)$ rapidly decreases to zero around  $r\sim r_0$. Therefore, if we have a relation like
\bea
\sqrt{h_0} \simeq {r_s \over l_s} ,
\label{sqrth0}
\eea
where $r_s$ is the Schwarzschild radius, the entropy \eqref{stringyEntropy} becomes $S \simeq r_s M $, which coincides with the entropy of the Schwarzschild black hole up to a numerical factor.

Although it would be difficult to show \eqref{sqrth0} rigorously, here we present a possible scenario to get it.
We assume that the system has no horizon, but the system chooses $r_0$ very close to $r_s$ in order to maximize the entropy,
\bea
r_0 = r_s +\delta, \qquad \delta \ll r_s .
\label{R}
\eea
The density changes from some finite value to zero as $r$ varies from $r_s$ to $r_0$. It is natural to imagine that such change occurs in the string scale,
which means that the physical distance between $r_s$ and $r_0$ is of order of the string scale $l_s$, that is,
\bea
\int_{r_s}^{r_0} dr \sqrt{h_0} \simeq \delta \sqrt{h_0} \simeq l_s.
\label{deltasqrth}
\eea
On the other hand, if we evaluate \eqref{h} around $r\simeq r_0$, we have
\bea
h_0 \simeq  \left( 1-{r_s \over r_0} \right)^{-1} \simeq {r_s  \over \delta} .
\label{hatr0}
\eea
Then \eqref{sqrth0} follows from \eqref{deltasqrth} and \eqref{hatr0}.

To summarize, in this paper we have studied systems consisting of strings, membranes and charges in the Einstein-Maxwell theory in $3+1$ dimensions.
We constructed the string hedgehog solution, which has, though charged, the
same causal structure as the Schwarzschild black hole (and thus has a single
horizon), whereas the singularity is replaced by a conical one.
Then we studied the gravitational charged membrane solution. We gave a
simple derivation of the self-energy of the membrane. The gravitational binding
energy that reduces the mass leads to the existence of unstable solutions as
well as stable ones.
We found that for each fixed value of the membrane tension there is a
maximal charge (and mass) where the solution approaches the extremal black hole,
while the interior of the membrane is flat Minkowski spacetime.
Finally, we constructed solutions by combining these two configurations to
show there are black hole solutions that have no singularities inside the
horizons.
We studied in detail the behavior of the solutions by varying the magnitude
of the charge (times the membrane tension), which is the only parameter of the
solutions if we fix the string tension.

There are several directions to generalize the solutions we have studied in this paper. For example, it is interesting to include the effect of rotations to see the relation to the Kerr-Newman solution.
It is also interesting to allow the radial motion to study dynamical nature of the models.
To generalize the present arguments to higher dimensions might also be interesting.
We hope to report some of these issues in the near future.

\section*{Acknowledgements}
The authors would like to thank K. Murakami, M. Ninomiya, Y. Sekino and F. Sugino for valuable discussions and comments.
T.M. thanks the particle theory group of Department of Physics at Kyoto University for hospitality during the completion of this work. This work was supported by the Grant-in-Aid for the Global COE Program "The Next Generation of Physics, Spun from Universality and Emergence" from the Ministry of Education, Culture, Sports, Science and Technology (MEXT) of Japan.


%

%
\appendix

\section{The energy-momentum tensors}
\label{EMT}
In this appendix we give the energy-momentum tensors for the Nambu-Goto $p$-branes.
The action of the $p$-brane is given by
\bea
S_{p}= -\tau_p \int d^{p+1} \sigma \sqrt{-\gamma} .
\eea
Here $\gamma$ is the determinant of the induced metric $\gamma_{ab}$ given by
\bea
\gamma_{ab}=\p_aX^\mu \p_b X^\nu g_{\mu\nu}(X) ,
\eea
where $g_{\mu\nu}$ is the spacetime metric, and $a,b = 0,1, \ldots, p$ and $\mu, \nu = 0,1, \ldots , d$. 
The energy-momentum tensor is in general defined by the functional derivative of the action with respect to the metric, 
\bea
\delta S = {1\over2} \int d^{d+1}x\sqrt{-g} T^{\mu \nu} \delta g_{\mu\nu} .
\eea
The energy-momentum tensor of the $p$-brane is thus given by
\bea
T^{\mu\nu}(x)
=-\tau_p \int d^{p+1} \sigma {\delta^{d+1}(x-X(\sigma)) \over \sqrt{-g(x)}} \sqrt{-\gamma} \gamma^{ab}\p_bX^\mu \p_b X^\nu .
\label{emt}
\eea
Below we consider spherically symmetric static geometries in $3+1$ dimensions.

\subsection{string}
\label{emtstring}
We consider a static string that extends radially from the origin in a static spherically symmetric spacetime.
We use spherical coordinates,
\bea
x^0=t,\quad x^1=r,\quad x^2=\theta ,\quad x^3=\varphi , 
\label{sphericalCoord}
\eea
and take the static gauge,
\bea
\sigma^0=X^0, \quad \sigma^1=X^1 . 
\eea
Then the energy-momentum tensor \eqref{emt} can be easily evaluated as
\bea
&&{T^{0}}_0(x) =-\mu {\delta(\theta-X^2)\delta(\varphi-X^3) \over \sqrt{g_{\theta \theta}g_{\varphi \varphi}}} = -\rho ,
\quad
{T^{1}}_1(x) = -\mu{\delta(\theta-X^2)\delta(\varphi-X^3) \over \sqrt{g_{\theta \theta}g_{\varphi \varphi}}}  =P_r ,
\nonumber \\
&& 
{T^{2}}_2(x) ={T^{3}}_3(x) =0=P_\perp .
\eea
We are interested in string hedgehog consisting of $N$ strings with a uniform angular distribution. 
By taking the angular average, we obtain 
\bea
&&{T^{0}}_0(x) =-{N\mu  \over 4\pi r^2}= -\rho ,
\quad
{T^{1}}_1(x) = -{N\mu  \over 4\pi r^2}  =P_r ,
\nonumber \\
&& 
{T^{2}}_2(x) ={T^{3}}_3(x) =0=P_\perp .
\label{emtNstrings}
\eea

\subsection{membrane}
\label{emtmemb}

We consider a static spherical membrane located at $r=r_0$ in a static spherically symmetric spacetime.
We use the spherical coordinates \eqref{sphericalCoord}, and take the static gauge,
\bea
\sigma^0=X^0, \quad \sigma^1=X^2,  \quad \sigma^2=X^3 .
\eea
Then the energy-momentum tensor \eqref{emt} is evaluated as
\bea
{T^{0}}_0(x) =-\kappa {\delta(r-r_0) \over \sqrt{g_{rr}}} = -\rho ,
\eea
and
\bea
{T^{1}}_1(x) =0=P_r, \quad
{T^{2}}_2(x) ={T^{3}}_3(x) = -\kappa {\delta(r-r_0) \over \sqrt{g_{rr}}} =P_\perp .
\eea

\section{Minimization of ADM mass and equation of motion}
\label{eom}
In this appendix we explicitly show the equivalence of the minimization of the ADM mass and the equilibrium equation.

We start with the equilibrium equation \eqref{balance},
\bea
{d P_r \over dr}+{2 \over r} \left( P_r - P_\perp \right) +{h \over r^2}\left(m+4\pi r^3 P_r\right) \left(\rho +P_r \right) = 0 ,
\nonumber
\eea
where $\rho$, $P_r$ and $P_\perp$ are given by \eqref{dmemb}, \eqref{prmemb} and \eqref{ppmemb}, respectively.
We multiply $\sqrt{h}$ and integrate over a small region from $r_0-\epsilon$ to $r_0+\epsilon$. Dropping the terms that do not contain delta function, we have
\bea
\int_{r_0-\epsilon}^{r_0+\epsilon} dr \sqrt{h} {d P_r \over dr} 
+{2\kappa \over r_0} 
+\int_{r_0-\epsilon}^{r_0+\epsilon} dr
{h^{3/2} \over 4\pi r^4}\left(m+4\pi r^3 P_r\right) {dm \over dr}
=0 ,
\label{integeq}
\eea
where we have used $dm/dr=4\pi r^2 \rho$. 
In the small $\epsilon$ limit, we can replace $r$ with $r_0$ in the integrand. Then the first and the last terms combine to give a total derivative,
\bea
\int_{r_0-\epsilon}^{r_0+\epsilon} dr
\left[
\sqrt{h} {d P_r \over dr} +{h^{3/2} \over r_0} {dm \over dr} P_r 
\right]
&=&
\left[
\sqrt{h} P_r
\right]_{r_0-\epsilon}^{r_0+\epsilon}
\nonumber \\
&=&
-{Q^2 \over 8\pi r_0^4} {1 \over 1-4\pi \kappa r_0} ,
\label{firstEQ}
\eea
where we have used 
\bea
\sqrt{h(r_0+\epsilon)}={1 \over \sqrt{1-{2 m_0 \over r_0}}} ={1\over 1-4\pi \kappa r_0} ,
\eea
where $m_0 =m(r_0+\epsilon)= 4\pi \kappa r_0^2(1-2\pi \kappa r_0)$.
Here we have assumed 
$4\pi \kappa r_0 <1$, which means $\tilde{r_0} <1$ and thus the stability of the membrane. 
The remaining integral can be calculated similarly as
\bea
\int_{r_0-\epsilon}^{r_0+\epsilon} dr {h^{3/2} \over 4\pi r^4}m {dm \over dr} 
&=&
\int_{m(r_0-\epsilon)=0}^{m(r_0+\epsilon)=m_0} dm {m \over 4\pi r_0^4 \left(1- {2m \over r_0}\right)^{3/2}}
\nonumber \\
&=&
{2\pi \kappa ^2 \over 1-4\pi \kappa r_0} .
\label{secondEQ}
\eea
Using \eqref{firstEQ} and \eqref{secondEQ}, we find that \eqref{integeq} becomes
\bea
{1 \over 4\pi r_0^2(1-4\pi \kappa r_0)}
\left[
8\pi \kappa r_0 - 24 \pi^2 \kappa^2 r_0^2-{Q^2 \over 2 r_0^2}
\right]
=0 ,
\eea
which is equivalent to the minimization of the ADM mass, 
\bea
{dm_\infty \over dr_0}
=0 ,
\eea
where $m_\infty$ is given by \eqref{minfCM0}.

\end{document}